\documentclass[11pt]{article}
\usepackage{epsfig}
\usepackage{cite}
\usepackage{color}

\definecolor{red}{named}{Red}

\newcommand{\be}{\begin{equation}}
\newcommand{\ee}{\end{equation}}
\newcommand{\beeq}{\begin{eqnarray}}
\newcommand{\eeeq}{\end{eqnarray}}

\def\bea{\begin{eqnarray}}
\def\eea{\end{eqnarray}}

\def\funp{{I\!\!P}}
\def\xp{x_{{I\!\!P}}}

\def\qbar{\overline{q}}

\def\gev{\mbox{\rm GeV}}

\def\eto{{\rm e}}

\newcommand{\eq}{\!\!\!&=&\!\!\!}
\def\half{{\textstyle{\frac{1}{2}}}}
\newlength{\dinwidth}
\newlength{\dinmargin}
\title{ \vspace {2.0cm}\bf
Diffractive parton distributions from the analysis with higher twist\vspace {0.5cm}}
\author{
Krzysztof~Golec-Biernat$^{(a,b)}$\footnote{e-mail: golec@ifj.edu.pl}~~and~ 
Agnieszka~\L{}uszczak$^{(b)}$\footnote{e-mail: agnieszka.luszczak@ifj.edu.pl}\\  \\
\textit{$^{a}$Institute of Nuclear Physics Polish Academy of Sciences, 
        Cracow, Poland}\\
\centerline{and}\\
\textit{$^{b}$Institute of Physics, University of Rzesz\'ow, Rzesz\'ow, Poland}\\ \\
}
\date{}

\begin{document}
\maketitle

\vspace*{0.5in}

\centerline{(\today)}

\vspace*{0.5in}

\begin{abstract}
\noindent

Diffractive parton distributions of the proton are determined from fits to  diffractive data
from HERA. In addition to the twist--2 contribution, the twist--4 contribution from longitudinally polarised virtual photons is considered, which is  important in the region
of small diffractive masses. A new prediction for the longitudinal diffractive structure function is  presented which differs significantly from that obtained in 
the pure twist--2 analyses.

\end{abstract}

\newpage
\section{Introduction}

The diffractive deep inelastic scattering (DDIS) at HERA provide
a very interesting example of the interplay between 
hard and soft aspects of QCD interactions.
On one side, the virtuality of the photon  probe is large
($Q^2\gg \Lambda_{QCD}^2$), while  on the other side, the scattered proton remains almost intact,
loosing only a small fraction of its initial momentum. Its  
transverse momentum  with respect to the photon-proton collision axis is  also small.  
In addition to the scattered incident particles,  a diffractive system forms which is well separated in rapidity from the scattered proton.
The most important observation made at HERA is that diffractive processes
in DIS are not rare, quite the contrary, they constitute up to $15\%$ of  deep inelastic events.
What's more, the ratio of the diffractive and inclusive cross sections is constant as a function of energy of the $\gamma^* p$ system  or as a function of
the photon virtuality. The latter fact reflects the logarithmic dependence on $Q^2$
of diffractive structure functions in the Bjorken limit.

In the $t$-channel picture, the diffractive interactions can be viewed as
a vacuum quantum number exchange between the diffractive system and the proton. 
In old days of Regge phenomenology  such a {\it mechanism} of interactions was termed a {\it pomeron}.
With the advent of quantum chromodynamic we gain a new way of understanding the pomeron by modelling it with the help
of gluon exchanges projected onto the color singlet state. In the lowest approximation, the pomeron is a two gluon exchange which is independent of energy. By considering radiative corrections to this process  in the high energy limit, the famous BFKL pomeron \cite{Fadin:1975cb,Lipatov:1976zz,Kuraev:1977fs,Balitsky:1978ic} was constructed with a strong, power-like dependence on energy. This dependence ultimately  violates unitarity
which means that exchanges with more gluons have to be considered. 
A systematic program to sum  exchanges with gluon number changing vertices 
was formulated in \cite{Bartels:1978fc,Bartels:1980pe} and developed in
\cite{Bartels:1992ym,Bartels:1993ih,Bartels:1994jj,Bartels:1999aw}. Other, somewhat more
intuitive formulation, called Color Glass Condensate 
\cite{McLerran:1993ni,McLerran:1993ka,Iancu:2003xm,Iancu:2006qi },
is based on the idea of parton saturation \cite{Gribov:1984tu}
in which deep inelastic scattering occurs on a dense gluonic system in the proton. In these approaches unitarization is supposed 
to change the asymptotic energy behaviour of the cross sections involving the pomeron from power-like to logarithmic. 

DDIS  is  particularly sensitive to the pomeron energy behaviour  
since diffractive scattering amplitudes are squared in diffractive cross sections.  
Thus, unitarization effects
play more important  role than for the total cross section
which is proportional to the imaginary part of the scattering amplitude.
This observation is a basis of  a successful description of the first diffractive data from
HERA in which the diffractive system was formed by the quark-antiquark ($q\qbar$) and
quark-antiquark-gluon  ($q\qbar g$) systems. They  can be viewed in the space
of Fourier transformed transverse momenta as color dipoles \cite{Nikolaev:1990ja, Nikolaev:1991et}. 
In the approach  we follow in the forthcoming presentation, the pomeron interaction is modelled by a two-gluon exchange  which is subsequently substituted by the effective dipole--proton cross section 
fitted to inclusive DIS data \cite{Golec-Biernat:1998js,Golec-Biernat:1999qd}. In this way unitary is achieved.

In an alternative approach to DDIS, the diffractive structure functions are defined in terms
of diffractive parton distributions (DPD). They are evolved in $Q^2$ with the help of the 
Dokshitzer-Gribov-Lipatov-Altarelli-Parisi (DGLAP) equations \cite{Gribov:1972ri,Altarelli:1977zs,Dokshitzer:1977sg}. 
Thus in the Bjorken limit, the diffractive structure functions depend logarithmically on $Q^2$,
i.e. they provide the twist--2 description of DDIS. 
The theoretical justification of this approach is provided by the collinear factorisation theorem which is valid for hard diffractive scattering in $ep$ collisions \cite{Berera:1995fj,Collins:1997sr,Hautmann:1998xn,Hautmann:1999ui,Blumlein:2006ia}.
However, collinear factorisation fails in hadron--hadron scattering  
due  to nonfactorizable soft interactions between incident hadrons \cite{Collins:1992cv,Wusthoff:1999cr}.
Thus, unlike inclusive parton distributions, the DPD are  not universal objects 
and in general can only be used  for diffractive processes in the $ep$ deep inelastic scattering. Nevertheless, the scale of nonuniversality can be estimated by applying them
to hadronic reactions.

The relation between the color dipole approach with the $q\qbar$ and $q\qbar g$ diffractive components and the DGLAP based description was studied in detail in  \cite{Golec-Biernat:2001mm}. In short, after extracting the twist--2 part, the dipole
approach provides $Q^2$-independent quark and gluon DPD.  In addition, the  $q\qbar g$ 
component,  which was computed assuming strong ordering between transverse momenta
of the gluon and the $q\qbar$ pair, gives the first step in the $Q^2$-evolution
of the gluon distribution. 
The twist--2 approach, which is based on the DGLAP evolution equations, extends the  
two component dipole picture by taking into account more complicated diffractive
final state. In the performed up till now twist-2 analyses,
the diffractive parton distributions are determined through
fits to diffractive data from HERA \cite{Aktas:2006hy} 
We will follow this approach with an important modification.

The dipole approach teaches us one important lesson concerning 
the seemingly subleading twist--4 contribution, given by the $q\qbar$ pair from longitudinally polarised virtual photons $(Lq\qbar)$. Formally, it is  is suppressed by a  power of $1/Q^2$ with respect to the  leading twist--2 transverse contribution. However, the perturbative QCD calculation shows that for small diffractive masses, $M^2\ll Q^2$, the longitudinal contribution dominates over 
the twist--2 one which tends to zero in this limit.
The effect of the $Lq\qbar$ component is particularly important
for the longitudinal diffractive structure function $F_L^D$ which is supposed to be determined from the high luminosity run data at HERA. Thus, we claim that it is 
absolutely necessary to consider the twist--4 contribution in the determination of the diffractive parton distributions. The analysis which we present confirms its relevance  for  the prediction for $F_L^D$, which  
differs significantly from that based on the pure DGLAP analysis.
This is the main result of our paper.

The paper is organised as follows. In Section~\ref{sec:2} we provide basic formulae
for the kinematical variables and quantities measured in diffractive deep inelastic scattering. We also describe the 
three contributions which we include in the description of the diffractive structure functions, i.e.
the twist--2, twist--4 and Regge contributions. In Section~\ref{sec:3} we describe
performed fits while 
in Section~\ref{sec:4} we present their impact on the determination of the diffractive parton distributions  and diffractive structure functions.
We finish with conclusions and outlook.

\begin{figure}[t]
\begin{center}
\epsfig{figure=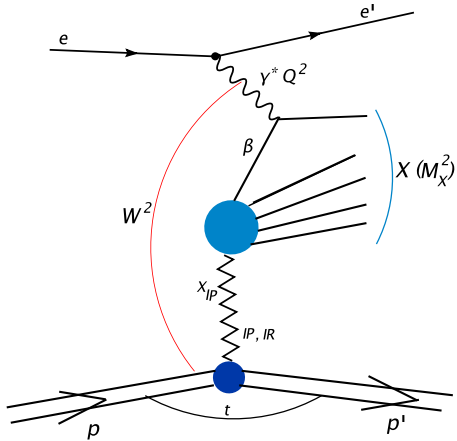,width=7cm}
\caption{Kinematic variables relevant for diffractive DIS.}
\label{fig:00}
\end{center}
\end{figure}

\section{Basic formulae}
\label{sec:2}

We consider diffractive deep inelastic scattering: $ep\to e^\prime p^\prime X$, 
shown schematically in  Fig.~\ref{fig:00}. After averaging over the azimuthal angle of the scattered proton,  
the four-fold differential cross section is given in terms of the diffractive
structure functions $F_2^D$ and $F_L^D$:
\be\label{eq:1}
\frac{d^4\sigma^D}{d\beta\, dQ^2\, dx_\funp\, dt}=\frac{2\pi\alpha^2_{em}}{\beta\, Q^4}
\left(1+(1-y)^2\right)
\left\{F_2^D-\frac{y^2}{1+(1-y)^2}F_L^D\right\}
\ee
where $y={Q^2}/(x_B s)$ and  ${s}$ is the $ep$ centre-of-mass energy squared.
The expression in the curly bracket is called reduced cross section:
\be\label{eq:2}
\sigma_r^{D}=F_2^D-\frac{y^2}{1+(1-y)^2}\,F_L^D\,.
\ee
Both structure functions depend on four kinematic variables $(\beta,Q^2,x_\funp,t)$,
defined as follows
\be\label{eq:3}
x_\funp=\frac{Q^2+M^2-t}{Q^2+W^2}\,,~~~~~~~~~~~~~~~~~~\beta=\frac{Q^2}{Q^2+M^2-t}\,,
\ee
where $-Q^2$  is virtuality of the photon, $t=(p-p^\prime)^2<0$ is the square of
four-momentum transferred  into the diffractive system, $M$ is invariant diffractive mass and $W$ is invariant energy of the $\gamma^*p$ system. The Bjorken variable $x_B=x_\funp\,\beta$.
For most of the diffractive events $|t|$ is much smaller then other scales, thus
it can be neglected in eqs.~(\ref{eq:3}).
The diffractive structure functions are measured in a limited range of $t$, thus the integrated structure functions are defined
\be\label{eq:4}
F^{D(3)}_{2,L}(\beta,Q^2,\xp)=\int_{t_{min}}^{t_{max}} dt\,
F_{2,L}^D(\beta,Q^2,x_\funp,t)\,,
\ee
The integrated reduced cross section $\sigma_r^{D(3)}$ is defined in a similar way.

\subsection{Twist--2 contribution}

In the QCD approach based on collinear factorisation, the diffractive  struc\-ture functions are decomposed into the leading  and higher twist contributions
\be\label{eq:5}
F_{2,L}^D\,=\,F_{2,L}^{D({tw}2)}
\,+\,{F_{2,L}^{D({tw}4)}}\,+\,\ldots\,.
\ee
The twist--2 part is given in terms
of the diffractive parton distributions through the standard
collinear factorisation formulae
\cite{Trentadue:1993ka,Berera:1994xh,Collins:1994zv,Berera:1995fj}. 
In the next-to-leading logarithmic approximation we have
\beeq
\label{eq:6a}
F_{2}^{D({tw}2)}(x,Q^2,\xp,t) \eq 
S_D + \frac{\alpha_s}{2\pi}\left\{C^S_{2}\otimes S_D + C^G_{{2}}\otimes G_D\right\}
\\\nonumber
\\\label{eq:6b}
F_{L}^{D({tw}2)}(x,Q^2,\xp,t) \eq 
\frac{\alpha_s}{2\pi}\left\{C^S_{L}\otimes S_D + C^G_{{L}}\otimes G_D\right\}
\eeeq
where $\alpha_s$ is the strong coupling constant and $C_{2,L}^{S,G}$ are coefficients functions known from inclusive DIS \cite{Furmanski:1980cm,Furmanski:1981cw}.   The integral convolution is performed for the  longitudinal momentum fraction
\be
(C\otimes F)(\beta)= \int_\beta^1 dz\,  C\left({\beta}/{z}\right) F(z) \,.
\ee
Notice that in the leading order, when terms proportional to $\alpha_s$ are neglected, the longitudinal structure function $F_L^{D(tw2)}=0$.
The functions $S_D$ and $G_D$ are  given by diffractive quark 
and gluon distributions, $q_D^f$ and $g_D$:
\beeq\label{eq:7}
S_D \eq
\sum_{f=1}^{N_f} e_f^2\,\beta\left\{q^f_D(\beta,Q^2,\xp,t)+
\overline q^f_D(\beta,Q^2,\xp,t)\right\}
\\\nonumber
\\
G_D\eq\beta g_D(\beta,Q^2,\xp,t)
\eeeq
Note that $\beta={x}/{\xp}$ plays the role of the Bjorken variable in DDIS.
In the infinite momentum frame, the DPD are interpreted
as  conditional probabilities to find a parton
with the momentum fraction $x=\beta \xp$ in a proton under the condition that
the incoming proton stays intact losing
a small fraction $\xp$ of its  momentum.
A formal definition of the diffractive parton distributions based
on the quark and gluon twist-2 operators is given in \cite{Berera:1995fj,Hautmann:1998xn}.

The DPD are evolved in $Q^2$ by  the DGLAP evolution equations \cite{Collins:1998rz}
for which the variables $(\xp,t)$ play the role of  external parameters. In this analysis we
assume {\it Regge factorisation} for these variables:
\beeq
\label{eq:reggefac}
q^f_D(\beta,Q^2,\xp,t)\eq\,f_\funp(\xp,t)\,\,q_\funp^f(\beta,Q^2)
\\
g_D(\beta,Q^2,\xp,t)\eq\,f_\funp(\xp,t)\,\,g_\funp(\beta,Q^2)\,.
\eeeq
For convenience, the functions $q_\funp^f(\beta,Q^2)$ and $g_\funp(\beta,Q^2)$ are called pomeron parton distributions.
The motivation for such a factorisation is a model of diffractive interactions with a pomeron exchange
\cite{Ingelman:1984ns}. In this model $f_\funp$ is the pomeron flux
\be\label{eq:pomflux}
f_\funp(\xp,t)\,=\,\frac{F^2_\funp(t)}{8\pi^2}\,{\xp^{1-2\,\alpha_\funp(t)}}\,,
\ee
where $\alpha_\funp(t)=\alpha_{\funp}(0)+\alpha_{\funp}^\prime t$ is the pomeron  Regge trajectory and the formfactor
\be\label{eq:formfactor}
F^2_\funp(t)\,=\,F^2_\funp(0)\,\eto^{-B_D |t|}
\ee
describes the pomeron coupling to the proton. 
We set $F^2_\funp(0)=54.4~\gev^{-2}$ \cite{Collins:1994zv}, 
$B_D=5.5~\gev^{-2}$ and $\alpha_{\funp}^\prime=0.06~\gev^{-2}$ \cite{Aktas:2006hy},
while the pomeron intercept $\alpha_{\funp}(0)$ is fitted to data.
The pomeron quark  distributions are flavour independent, thus they 
are given by one function, a singlet quark distribution $\Sigma_\funp$:
\be
q_\funp^f(\beta,Q^2)\,=\,\overline{q}_\funp^f(\beta,Q^2)\,\equiv\,
\frac{1}{2N_f}\,\Sigma_\funp(\beta,Q^2)
\ee
where $N_f$ is a number of active flavours.
The question about Regge factorisation  is an issue which should be tested experimentally. 
In our approach,  the pomeron  is a model of diffractive interactions which only provides
energy dependence through the $\xp$-dependent pomeron flux.
Its normalisation is only a useful convention
since the normalisations of the pomeron parton distributions in eqs.~(\ref{eq:reggefac}) are fitted to data.

\subsection{Twist-2 charm contribution}

We describe the charm quark diffractive production using twist-2 formulae for the 
$c\overline{c}$  pair  generation  from a gluon. These are formulae analogous to the inclusive case in which the diffractive
gluon distribution $g^D$ is substituted for the inclusive one \cite{Gluck:1994uf}:
\be
\label{eq:fcb}
F_{2,L}^{D(c\overline{c})}(\beta,Q^2,\xp,t)\, =\, 
2\beta\, e_c^2\, \frac{\alpha_s(\mu_c^2)}{2 \pi}
\int_{a\beta}^1\frac{dz}{z}\,
C_{2,L}\!\left( \frac{\beta}{z},{m_c^2\over Q^2} \right) g^D(z,\mu_c^2,\xp,t)~ ,
\ee
where $a = 1 + 4 m_c^2/Q^2$ and the factorisation scale $\mu_c^2=4 m_c^2$
with the charm quark mass $m_c=1.4~\gev$.
The coefficient functions read
\beeq\nonumber
C_2(z, r) \eq \half\left\{z^2 + (1-z)^2 + 4z(1-3z) r -
8z^2 r^2\right\} \ln{1+\alpha\over 1 -\alpha} 
\\
&&~~+\, \half \alpha\left\{ -1 + 8z(1-z)-4z(1-z)r \right\}
\\ \nonumber
\\
C_L(z, r) \eq - 4z^2 r \ln{1+\alpha\over 1 -\alpha} + 2\alpha z (1 - z)
\eeeq
with  $\alpha = \sqrt{1 - {4 r z/(1-z)}}$. 
The $c\overline{c}$ pair can only be produced if invariant mass of the diffractive system
$M^2$ fulfils the following condition 
\be
M^2=Q^2 \left(\frac{1}{\beta} -1\right) >\, 4m_c^2\,.
\ee

\subsection{Twist--4 contribution}

\begin{figure}[t]
  \vspace*{-1.5cm}
     \centerline{
         \psfig{figure=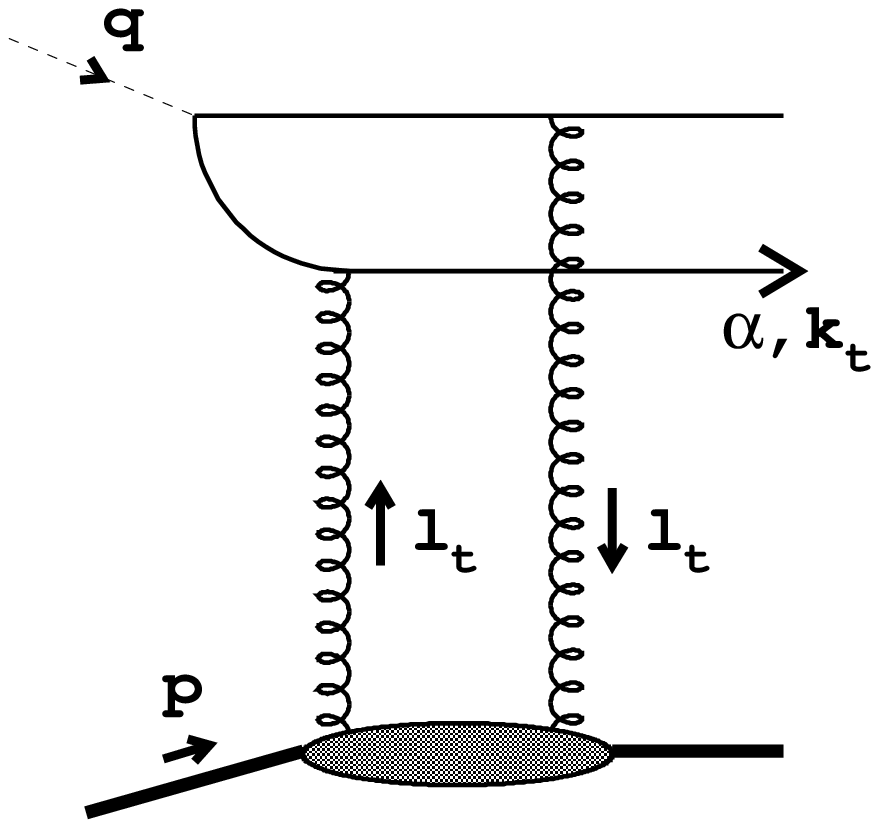,width=6cm}
         \psfig{figure=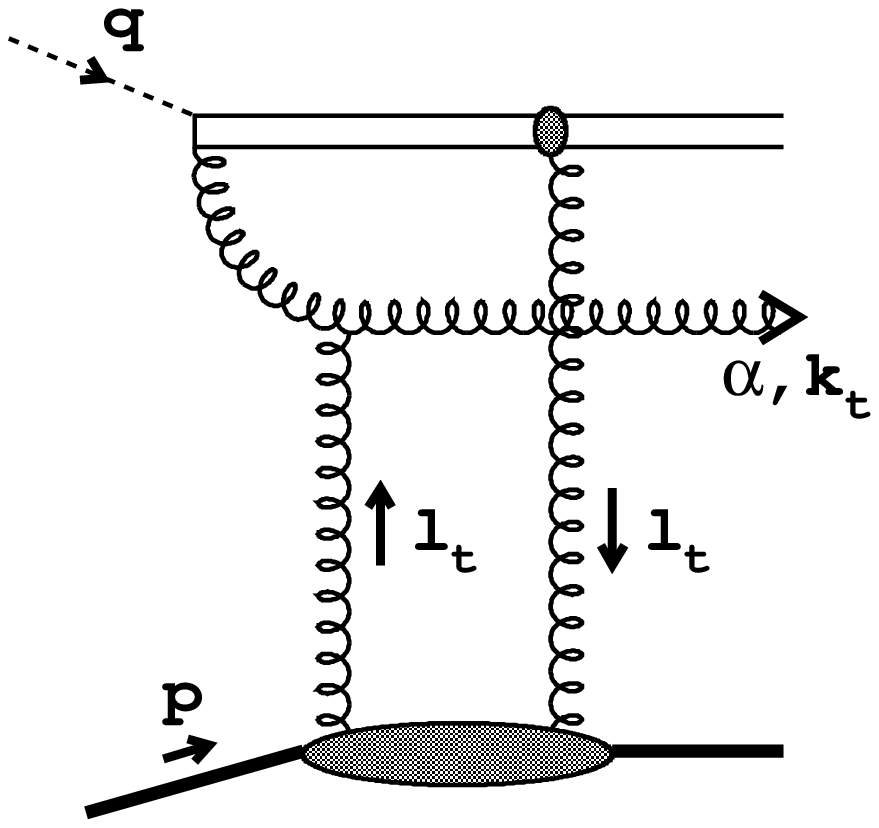,width=6cm}
           }
\vspace*{-1cm}
\caption{\it The $q\bar{q}$ and $q\bar{q}g$ components of the diffractive system in the dipole approach.
\label{fig:01}}
\end{figure}

The computation of the twist--4 contribution,  proportional to $1/Q^2$, 
is a nontrivial task and one could be tempted
to assume that this contribution is suppressed at large $Q^2$ as in inclusive DIS. 
However, by analysing diffractive final
states in the dipole approach it was found that for diffractive mass
$M^2\ll Q^2$ ($\beta\to 1$), the twist--4 contribution  dominates over the vanishing twist--2 one \cite{Wusthoff:1997fz,Bartels:1998ea,Golec-Biernat:1999qd}.

This observation is made on the basis of the perturbative QCD calculations in which
the diffractive state is formed by the $q\qbar$ and $q\qbar g$ systems interacting with
a proton through a colorless gluonic exchange which is a model of the pomeron interactions
in QCD. 
In the simplest case, two gluons projected onto the color singlet state are exchanged,
see Fig.~\ref{fig:01}. The computed amplitudes do not depend on energy in such a case
which problem can be cured in a more sophisticated approach by modelling 
the dipole-proton cross section which fulfils unitarity conditions  \cite{Golec-Biernat:1999qd}.
Independ of the details of the pomeron description, 
the diffractive mass (or $\beta$) dependence is a genuine prediction of pQCD calculations. It appears that the leading in $Q^2$ behaviour components, $q\qbar$ and $q\qbar g$ from 
transverse virtual photons, vanish for $\beta\to 1$.  This is not the case for the $q\qbar$ production from
longitudinal photons ($Lq\qbar$) which is formally suppressed by $1/Q^2$ with respect
to the leading components. Thus, this particular $\beta$-dependence makes the  $Lq\qbar$ contribution dominant for  $\beta\to 1$, see Fig.~\ref{fig:0}.

The  presence of the $Lq\qbar$ component has important consequence for the longitudinal
diffractive structure function which is supposed to be determined from the HERA data.
The formula given below is an important element in the description 
of $F_L^D$ in the region of large $\beta$:
\be
\label{eq:flqq}
F_{Lq\bar{q}}^{D}=
\frac{3}{16\pi^4\xp}\,\eto^{-B_D|t|}\,\sum_f e_f^2 \,
\frac{\beta^3}{(1-\beta)^4}\;
\int_0^{\frac{Q^2(1-\beta)}{4\,\beta}} \!dk^2\
\frac{\displaystyle {k^2}/{Q^2}}
{\displaystyle \sqrt{1-\frac{4\beta}{1-\beta}\frac{k^2}{Q^2}}}\,
\phi_0^2(k,\xp)
\ee
where the function $\phi_0(k,\xp)$ is given in terms of the dipole cross section
$\hat{\sigma}(\xp,r)$ and the Bessel functions $K_0$ and $J_0$:
\be
\label{eq:phi1}
\phi_0(k,\xp)
\;=\;k^2
\int_0^\infty dr\, r\, K_{0}\!\left(\sqrt{\frac{\beta}{1-\beta}}kr\right)
J_{0}(kr)\,  \hat{\sigma}(\xp,r)\,.
\ee
Strictly speaking, 
eq.~(\ref{eq:flqq})
contains all inverse powers of $Q^2$ but the part proportional to $1/Q^2$ (called
twist--4) dominates.
The dipole-proton cross section  describes the  interaction of 
a color dipole, formed by the $q\qbar$ or $q\qbar g$ systems,  with a proton. Following \cite{Golec-Biernat:1998js} we choose 
\be\label{eq:dipcs}
\hat{\sigma}(\xp,r)\,=\,\sigma_0\,\{1-\exp{(-r^2Q_s^2/4)}\}
\ee
where $Q_s^2=(\xp/x_0)^{-\lambda}~{\rm GeV}^2$ is a saturation scale which provides the energy dependence of the twist--4 contribution. 
The parameters
$\sigma_0=29~{\rm mb},~x_0=4\cdot10^{-5}$ and $\lambda=0.28$
are taken from  \cite{Golec-Biernat:1998js} (Fit 2 with charm).
This form of the dipole cross section provides  successful description of the first HERA data on both inclusive and diffractive structure functions \cite{Golec-Biernat:1998js,Golec-Biernat:1999qd}. A different  parametrisation
of $\hat{\sigma}$, without the saturation scale,  is also given in \cite{Forshaw:1999uf,Forshaw:1999ny,Forshaw:2004xd}.
We checked that a very similar description of $F_{Lq\bar{q}}^{D}$ was found
in a recent analysis \cite{Marquet:2007nf} based on the  Color Glass Condensate parameterisation of the dipole scattering amplitude \cite{Soyez:2007kg}.

The relation between the dipole approach with three diffractive components  and
the DGLAP approach with diffractive parton distributions was analysed at length in 
\cite{Golec-Biernat:2001mm}. Summarising this relation, the twist--2 part of 
the $q\qbar$ component gives a diffractive quark distribution.
The twist-2 part of the $q\qbar g$ component forms  a first step of the DGLAP evolution which starts from a given gluon distribution. Both diffractive parton distributions
do not depend on $Q^2$, thus they may serve as  initial conditions for the DGLAP equations at the scale which is not determined.
From this perspective,  the DGLAP approach offers a description of more complicated diffractive state with any number of partons ordered in transverse momenta.  However, the pQCD calculations tell us that the twist--2 analysis 
of diffractive data should include the twist--4 contribution since it  cannot be neglected at large $\beta$. This is the strategy which we follow in our analysis.

We also borrow from the dipole approach  a general form in $\beta$
of the initial  quark distribution which vanishes at the endpoints $\beta=0,1$ (see eq.~(\ref{eq:dpdfq}) in which $A_q$ and $C_q$ are  positive). A very important aspect of Regge factorisation (\ref{eq:reggefac}) can also be motivated by the dipole approach.
It is a consequence of geometric
scaling of the dipole cross section (\ref{eq:dipcs}) \cite{Golec-Biernat:2001mm,Stasto:2000er}.

\subsection{Reggeon contribution}

The diffractive data from the H1 collaboration for higher values of $\xp$ 
hints towards a contribution which decreases with energy. This effect can
be described by reggeon exchanges in addition to the rising with energy pomeron exchange.
Following \cite{Golec-Biernat:1996xj,Golec-Biernat:1997vy},
we consider  the dominant isoscalar $(f_2,\omega)$ reggeon exchanges which lead to the following contribution to $F_2^D$:
\be\label{eq:facreg}
F^{D(R)}_{2}\,=\,\sum_R\,f_R(\xp,t)\,F_R(\beta,Q^2)\,.
\ee
This contribution breaks Regge factorisation of the diffractive structure function, however, its presence is necessary for $\xp>0.01$ \cite{Aktas:2006hx}. 
The reggeon flux $f_R$ is given by the formula analogous to eq.~(\ref{eq:pomflux}) 
\be
\label{reflux}
f_R(\xp,t)\;=\;
\frac{F_R^2(0)}{8\pi^2}\,\eto^{-|t|/\Lambda^2_R}\,|\eta_R(t)|^2
\,\xp^{1-2\,\alpha_R(t)}\,,
\ee
where $\alpha_R(t)=0.5475+1\cdot t$ is the reggeon trajectory.
From the Regge phenomenology of hadronic reactions $\Lambda_R=0.65\,\mbox{\rm GeV}$  and the 
reggeon--proton couplings are given by \cite{Golec-Biernat:1997vy}:
$F_{f_2}^2(0)=194\, \gev^{-2}$ and $F_{\omega}^2(0)=52\,\gev^{-2}$.
The functions 
\be
|\eta_R(t)|^2=4\cos^2[\pi \alpha_R(t)/2]\,,~~~~~~~~~|\eta_R(t)|^2=4\sin^2[\pi \alpha_R(t)/2]
\ee
are signature factors for even ($f_2$) and odd ($\omega$) reggeons, respectively. 
We could also consider isovector reggeons $(a_2,\rho)$ but their 
couplings to the proton are much smaller and we neglect them.
Finally, the reggeon structure function $F_R$ is given by \cite{Golec-Biernat:1997vy}
\be
\label{eq:F_2R}
F_R(\beta) \,=\, A_R\, \beta^{-0.08}\,(1-\beta)^2\,,
\ee
where the normalisation $A_R$ is a fitted parameter. Thus, in the first approximation,
we neglect the $Q^2$-dependence of the reggeon contribution.

\section{Fit details}
\label{sec:3}

\begin{table}[ht]
\begin{center}
\begin{tabular}{|l||c|c|c|c|c|} 
\hline 
Collab. & No. points & Data & $|t|$-range   & $Q^2$-range 
& $\beta$-range\\
\hline\hline
H1~~~~ \cite{Aktas:2006hx}& 72   & LP  & $[0.08,\,0.5]$
 & $[2\,,50]$
& $[0.02\,,0.7]$ \\
\hline
ZEUS \cite{Chekanov:2004hy}&  80 & LP  &  $[0.075\,,0.35]$  & 
$[2\,,100]$ & $[0.007\,,0.48]$\\
\hline
H1~~~~ \cite{Aktas:2006hy}&  461  &  $M_Y< 1.6$  &   $[|t_{min}|\,,1]$  & $[3.5\,,1600]$ & $[0.01\,,0.9]$\\
\hline
ZEUS \cite{Chekanov:2005vv}&  198 &$M_Y< 2.3$  & $[|t_{min}|\,,\infty]$  & $[2.2\,,80]$ & $[0.003\,,0.975]$\\
\hline
\end{tabular}
\end{center}
\caption{Kinematic regions of diffractive data from HERA.
LP means leading proton data and $M_Y$ is invariant mass of
a  dissociated proton. Dimensionfull quantities are in units of $1~\gev$.}
\label{tabela:1}
\end{table}

In our analysis we use diffractive data from the H1 
\cite{Aktas:2006hx,Aktas:2006hy} and ZEUS \cite{Chekanov:2004hy,Chekanov:2005vv} collaborations.
In Table~\ref{tabela:1} we show their kinematic limits in which
they have been measured.
The minimal value value of $|t|$ is given by
\be\label{eq:mint}
|t_{min}|\simeq \frac{\xp^2}{1-\xp}\,m_p^2\,,
\ee
where $m_p$ is the proton mass. The leading proton data from H1, measured in the range
given  in Table~\ref{tabela:1}, were corrected by the H1 collaboration to the range $|t_{min}|<|t|<1~\gev^2$.

The ZEUS data are given for the diffractive structure function $F_2^D$, thus
we use in our analysis the following formulae
\beeq\label{eq:sf2la}
F_2^D\eq F_2^{D(tw2)}+\,F_{2}^{D(R)} +\,F_{Lq\bar{q}}^{D}
\\\label{eq:sf2lb}
F_L^D\eq F_L^{D(tw2)} +\, F_{Lq\bar{q}}^{D}\,.
\eeeq
The longitudinal twist-4 contribution is present on the r.h.s. of eq.~(\ref{eq:sf2la}) since  $F_2^D$ is the sum of the contributions from the transverse
and longitudinal polarised virtual photon.
The H1 data, however,  are presented for the reduced cross section (\ref{eq:2}). 
Thus we  substitute   relations (\ref{eq:sf2la}) and (\ref{eq:sf2lb}) in there and use
\be\label{eq:redcstw}
\sigma^D_r\,=\,\left\{F_2^{D(tw2)}+F^{D(R)}_2-\frac{y^2}{1+(1-y)^2}F_L^{D(tw2)}\right\}
\,+\,\frac{2(1-y)}{1+(1-y)^2}\,F_{Lq\qbar}^{D}\,.
\ee
The expression in the curly brackets  is the twist--2 contribution while the last term is
the twist--4 one. Notice that the difference between $F_2^D$ and $\sigma^D_r$  is most  important  for $y\to 1$.

We fit the diffractive parton distributions  at the initial scale
$Q_0^2=1.5~\gev^2$, assuming the Regge factorised form (\ref{eq:reggefac}) 
with the following pomeron parton distributions \cite{Aktas:2006hy}: 
\beeq\label{eq:dpdfq}
\beta\/\Sigma_\funp(\beta)\eq A_q\, \beta^{B_q}\,(1-\beta)^{C_q}
\\\label{eq:dpdfg}
\beta\/ g_\funp(\beta) \eq A_g\, \beta^{B_g}\,(1-\beta)^{C_g}\,.
\eeeq
The six indicated parameters are fitted to data.
We additionally multiplied both distributions by a factor $\exp\{-a/(1-\beta)\}$
with $a=0.01$ to secure their vanishing for $\beta=1$. This factor is only important
when $C_q$ or $C_g$ becomes negative in the fits. 
For the evolution, we use the next-to-leading order
DGLAP equations with $\Lambda_{QCD}=407~{\rm MeV}$ for $N_f=3$ flavours \cite{Martin:2006qz}.

The pomeron flux in eq.~(\ref{eq:reggefac})
is integrated over $t$ in the limits given in Table~\ref{tabela:1} which leads to the form
\be\label{eq:mintint}
f_\funp(\xp)\,=\,
\frac{F^2_\funp(0)}{8\pi^2B}
\left\{\eto^{-B|t_{min}|}-\,\eto^{-B|t_{max}|}\right\}\xp^{\,1-2\alpha_\funp(0)}\,.
\ee
The shrinkage parameter $B$ equals
\be
B=B_D+2\alpha_\funp^\prime\ln(1/\xp)
\ee 
with $B_D=5.5\,\gev^{-2}$ and $\alpha_\funp^\prime=0.06\,\gev^{-2}$ \cite{Aktas:2006hx}.

In summary, we have 
eight fit parameters altogether: the pomeron intercept $\alpha_\funp(0)$,  reggeon normalisation $A_R$ in eq.~(\ref{eq:F_2R})
and six parameters  in eqs.~(\ref{eq:dpdfq},\ref{eq:dpdfg})

\begin{table}[t]
\begin{center}
\begin{tabular}{|c||c||c||c|c||c|c|c||c|c|c||c|} 
\hline 
No& 
Data & Fit &
$\alpha_\funp(0)$ & $A_R$ &$A_q$  & $B_q$  & $C_q$  & $A_g$  & $B_g$  
&$C_g$&$\chi^2/N$\\
\hline\hline

1 &
H1 (LP) & tw-2 & 1.098& 0.29 & 1.75 &  1.49  & $0.5^*$ & 2.09 & 0.67 & 0.80  & 0.48 \\
\hline
2 &
ZEUS (LP) & tw-2 & 1.145& 1.05 & 2.13 &  1.51  & $0.5^*$ & 10.0* & 1.03 & 2.26  & 0.40\\
\hline\hline
3 &
H1  & tw-2 &  1.117 &  0.49 & 1.33  &  1.63 &  0.34 &  0.17  & -0.16 & -1.10 & 1.04
\\
\hline
4 &
& tw-(2+4)  & 1.119 & 0.48 & 1.62 & 1.98  & 0.59  & 0.04  & -0.56  & -1.68  & 1.17 \\
\hline\hline
5 &
ZEUS & tw-2 &  1.093 & $0.0^*$ & 1.68 & 1.01  & $0.5^*$ & 0.49 & -0.03 & -0.40 &  1.35 \\
\hline
6 &
    & tw-(2+4) &  1.092 & $0.0^*$ & 1.20 & 0.85  & 0.57 & 0.07 & -0.52  & -1.48 & 1.82
\\
\hline
\end{tabular}
\end{center}
\caption{The fit  parameters to H1 nd ZEUS data. The presence of twist--4
in the fits is marked by tw-(2+4).
The parameters with an asterisk are fixed in the fits.}
\label{tabela:2}
\end{table}

\section{Fit results}
\label{sec:4}

The data sets from Table~\ref{tabela:1} were obtained in different
kinematical regions,  using different methods of their analysis. Thus,
we decided to perform fits to each data set separately.
The  values of the fit parameter
are shown in  Table~\ref{tabela:2}. 
The difference between them can be attributed to the scale of uncertainty of our analysis. 
In each case we preformed two fits: 
with  and without the twist--4 formula added to the twist--2 contribution. 

\subsection{Leading proton data}

\begin{figure}[p]
\begin{center}
\psfig{figure=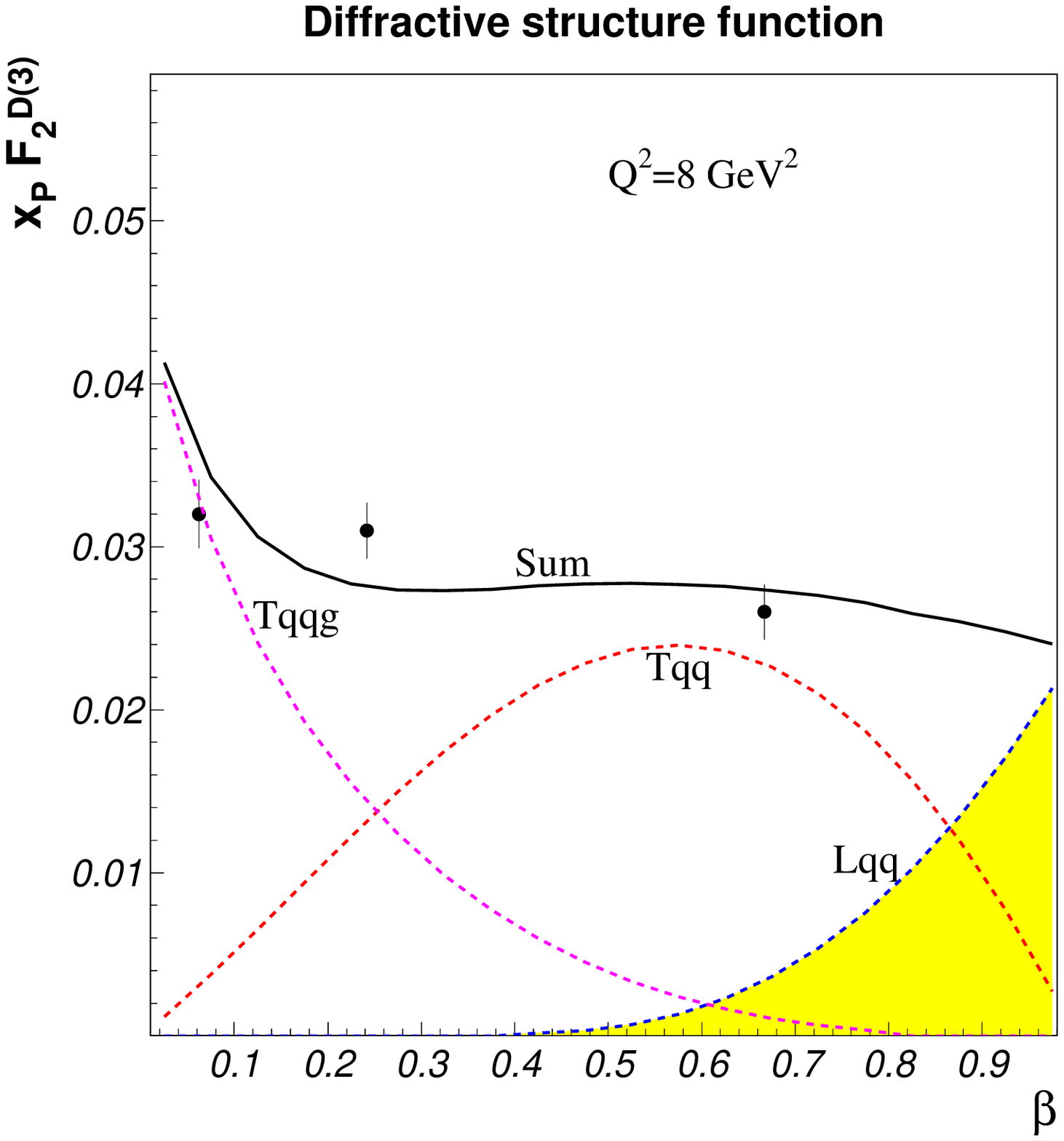,width=17cm}
\caption{Three contributions to $F_2^D$ from: $q\qbar$ and $q\qbar g$
from transverse (T) and longitudinal (L)  photons \cite{Golec-Biernat:1999qd}
for $\xp=0.003$.
The twist--4 contribution $Lq\bar{q}$ is indicated by the yellow
band. Old ZEUS data points are shown.}
\label{fig:0}
\end{center}
\end{figure}

We started from fits to the leading proton data. The fit parameters in this case are 
displayed in the first two rows of Table~\ref{tabela:2}. We only show  the twist--2 fit results
since they are not changed in fits with the twist--4 term.   
This happenes because the leading proton data comes from the  region of $\beta$ values
where the twist--4 contribution is  small ($\beta\le 0.7$ for H1 and  $\beta<0.5$ for ZEUS), see Fig.~\ref{fig:0}.

The data with a dissociated proton (DP) which are measured in the region of large $\beta$
influence most the value of  the parameter $C_g$  which controls the behaviour of the gluon distribution at $\beta\to 1$. 
For the LP data  $C_g$ is positive and the gluon distribution is suppressed near
$\beta\approx 1$, while for the DP data $C_g$ is negative  and the gluon distribution is strongly enhanced.
This shows that the data with $\beta >0.7$   are crucial for the proper analysis. 
Without this kinematic region  we lose important information about diffractive interactions.
Thus, from now on we concentrate on the analysis of the DP data.

\subsection{H1  data}

\begin{figure}[p]
\begin{center}
\psfig{figure=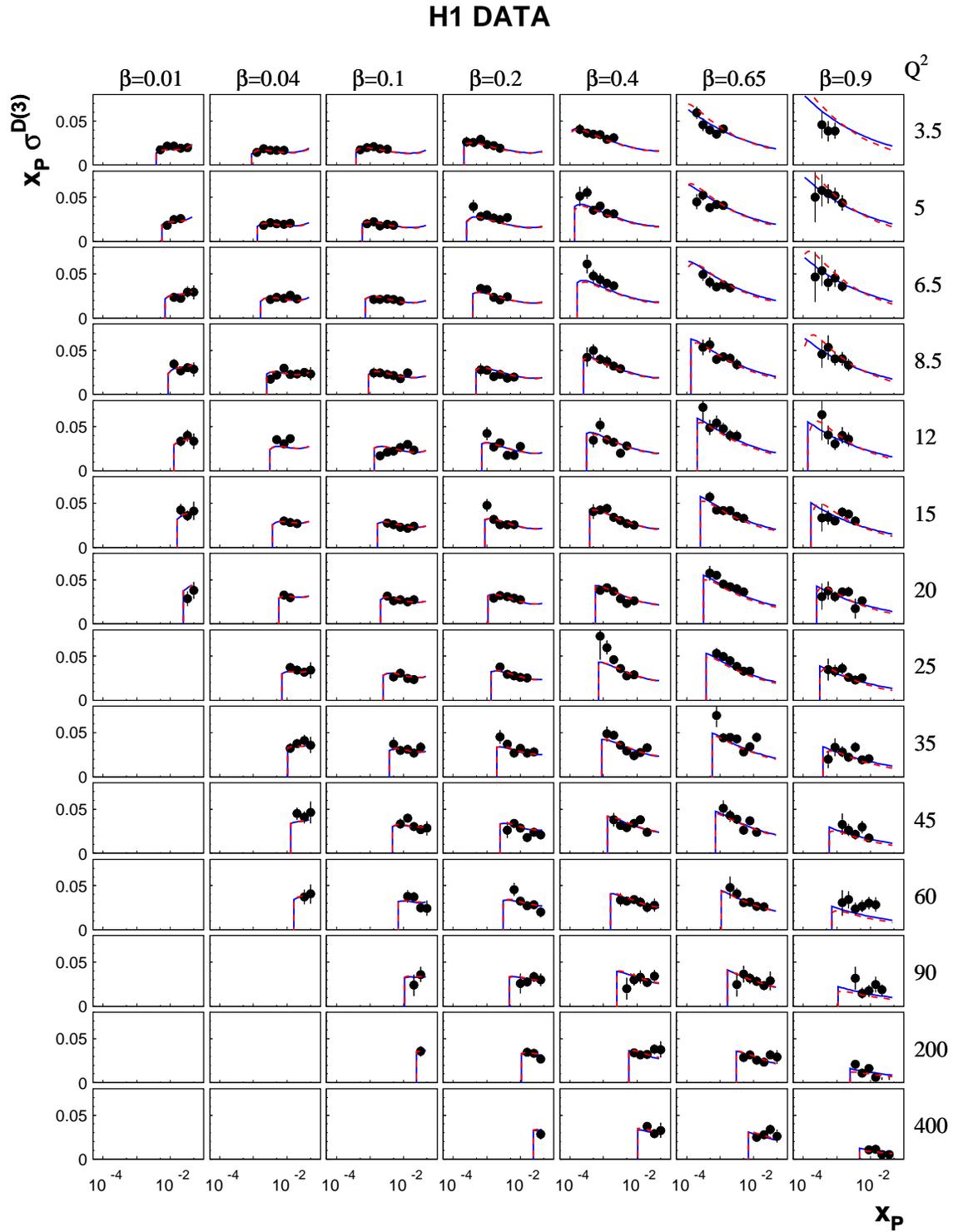,width=17cm}
\caption{Reduced cross section $\sigma_r^{D(3)}$ for H1 data as a function of $\xp$. Solid lines: twist--2 fit, dashed lines: twist--(2+4) fit.}
\label{fig:1}
\end{center}
\end{figure}

The fit parameters to the H1 data with a dissociate proton are given
in the third and fourth rows of Table~\ref{tabela:2}. We see that
the fit quality is practically the same for both fits, with and without the twist--4 contribution.
The presence of the reggeon term improves fit quality by 30 units of $\chi^2$ for
461 experimental points. A good quality of the fits is illustrated in Fig.~\ref{fig:1} 
which also shows that the reduced cross sections (\ref{eq:redcstw}) from the twist--2  
(solid lines) and twist--(2+4) fits (dashed lines) are very close to each other.

\begin{figure}[p]
\begin{center}
\psfig{figure=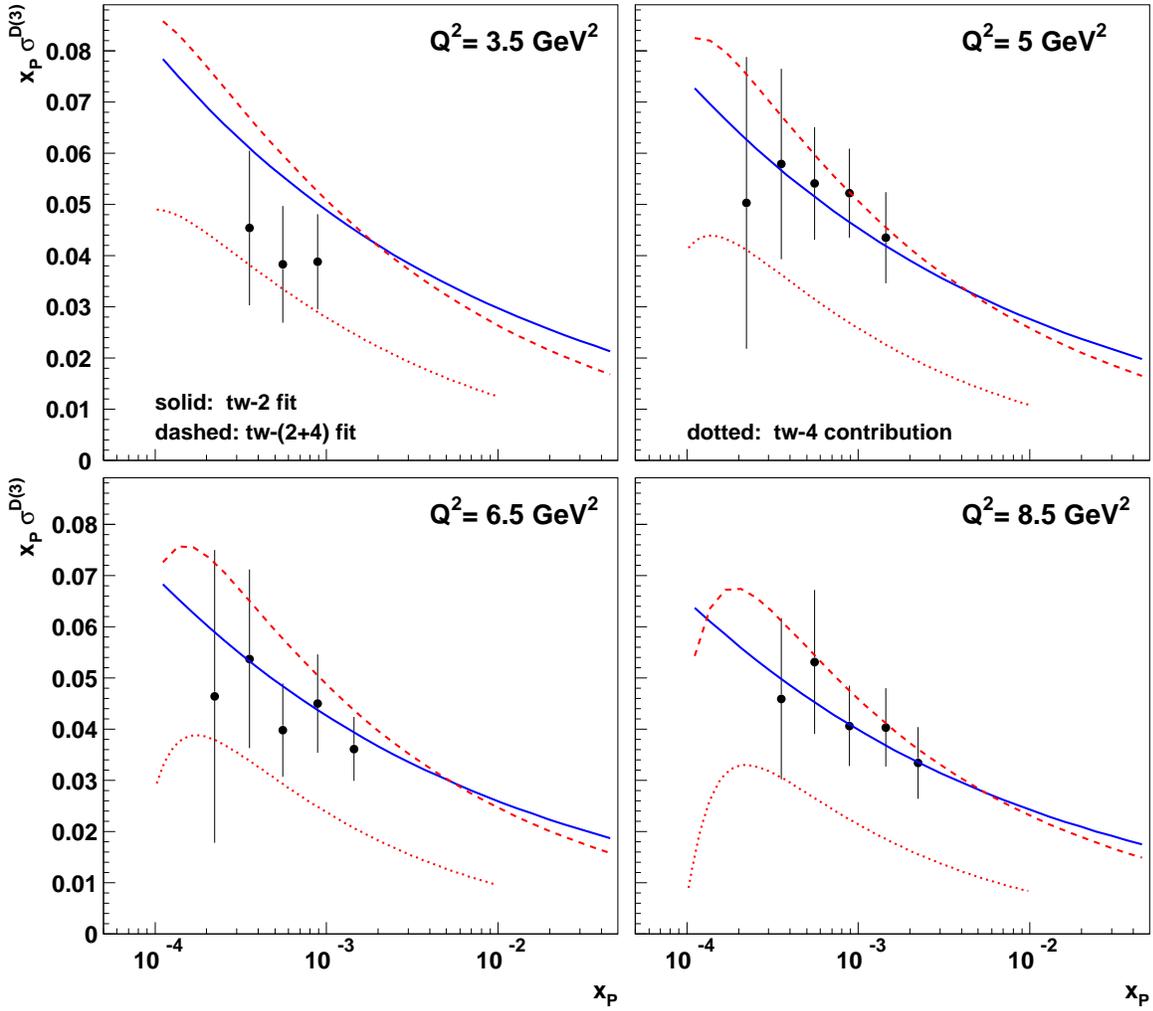,width=17cm}
\caption{Reduced cross section $\sigma_r^{D(3)}$ for H1 data at $\beta=0.9$ for four values
of $Q^2$ against fit curves.}
\label{fig:2}
\end{center}
\end{figure}

In Fig.~\ref{fig:2} we show our results on the reduced cross section
for the largest measured value of $\beta=0.9$. In this region,
the twist--4 contribution, shown by the dotted lines,  cannot be neglected. We see that 
the curves from both the twist--2  (solid) and twist--(2+4) (dashed) fits describe data reasonable well. However, the curves with twist--4  have a steeper dependence on $\xp$ (energy) than in the pure twist--2 analysis. This observation  is by far more pronounced in the analysis of the ZEUS data performed for the structure function $F_2^D$.

\begin{figure}[p]
\begin{center}
\psfig{figure=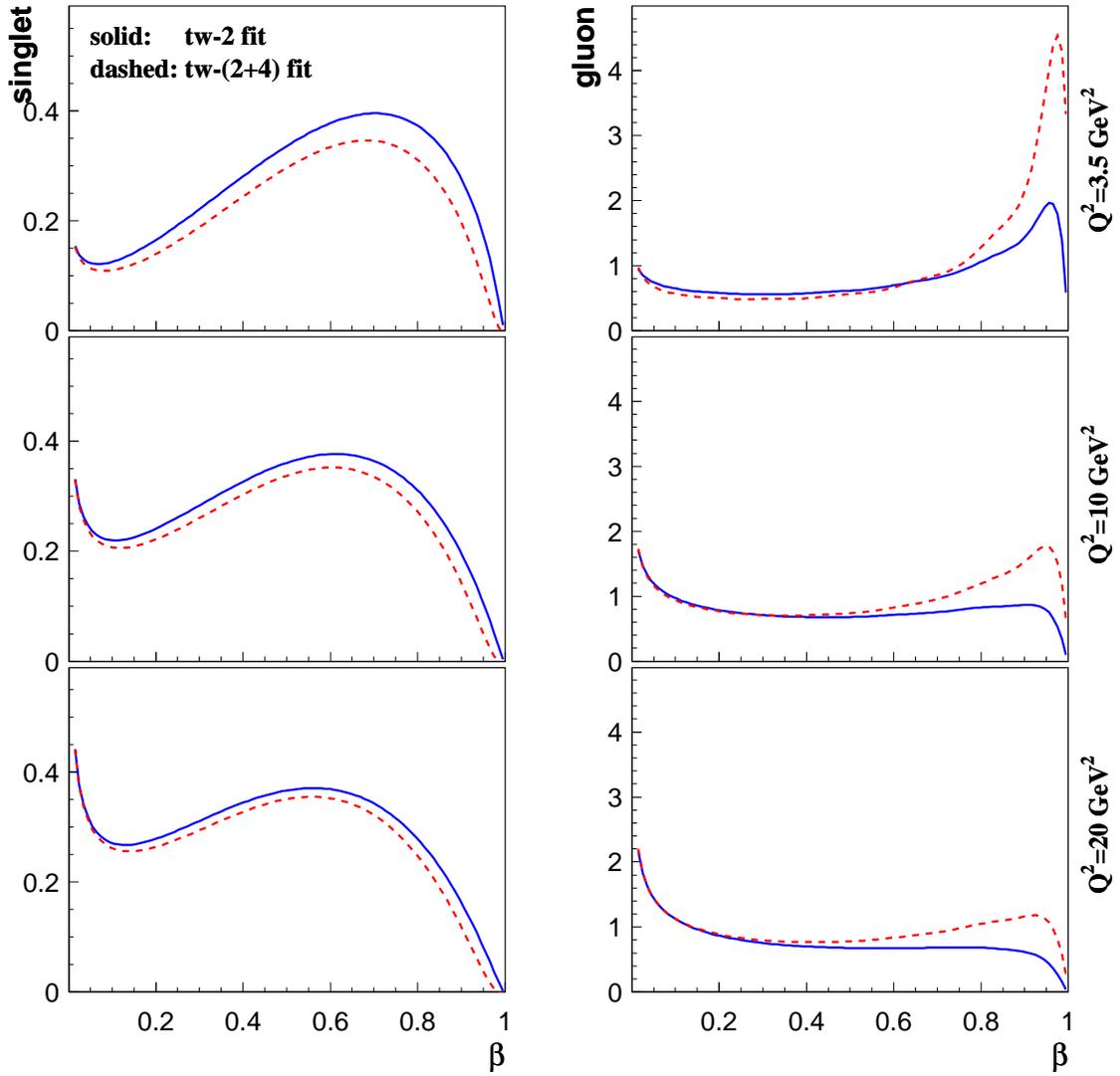,width=17cm}
\caption{Pomeron parton distributions: singlet $\beta\Sigma_{\funp}(\beta,Q^2)$ (left) and gluon  $\beta g_{\funp}(\beta,Q^2)$ (right) from H1 data.}
\label{fig:3}
\end{center}
\end{figure}

The diffractive parton distributions  from our fits
are shown in  Fig.~\ref{fig:3} in terms of the pomeron parton distributions,
$\beta \Sigma_{\funp}(\beta,Q^2)$ and $\beta g_{\funp}(\beta,Q^2)$. 
Being independent of the pomeron flux, such a presentation allows
for a direct comparison  of the results  from fits to different data sets.
We see   that the singlet quark distributions
are quite similar while the gluon distributions are different. 
In the fit with twist--4, the gluon distribution is stronger peaked near $\beta\approx 1$.
This somewhat surprising result can be understood by looking at the logarithmic slope
of $F_2^D$ for fixed values of $\beta$. From the LO DGLAP equations we have schematically:
\be
\frac{\partial F_2^D}{\partial \ln Q^2}\,\sim\,\frac{\partial\Sigma_\funp}{\partial \ln Q^2}
\,=\,P_{qq}\otimes \Sigma_\funp \,+\, P_{qG}\otimes G_\funp\,-\,\Sigma_\funp \int P_{qq}
\ee
where the negative term describes virtual corrections. For large $\beta$, the measured slope is negative which means that the virtual emission term must dominate over the real emission ones.
The addition of the  twist--4 contribution to $F_2^D$, proportional to $1/Q^2$, contributes a negative value to the slope which has to be compensated by a larger gluon distribution
in order to describe the same data.

\begin{figure}[p]
\begin{center}
\psfig{figure=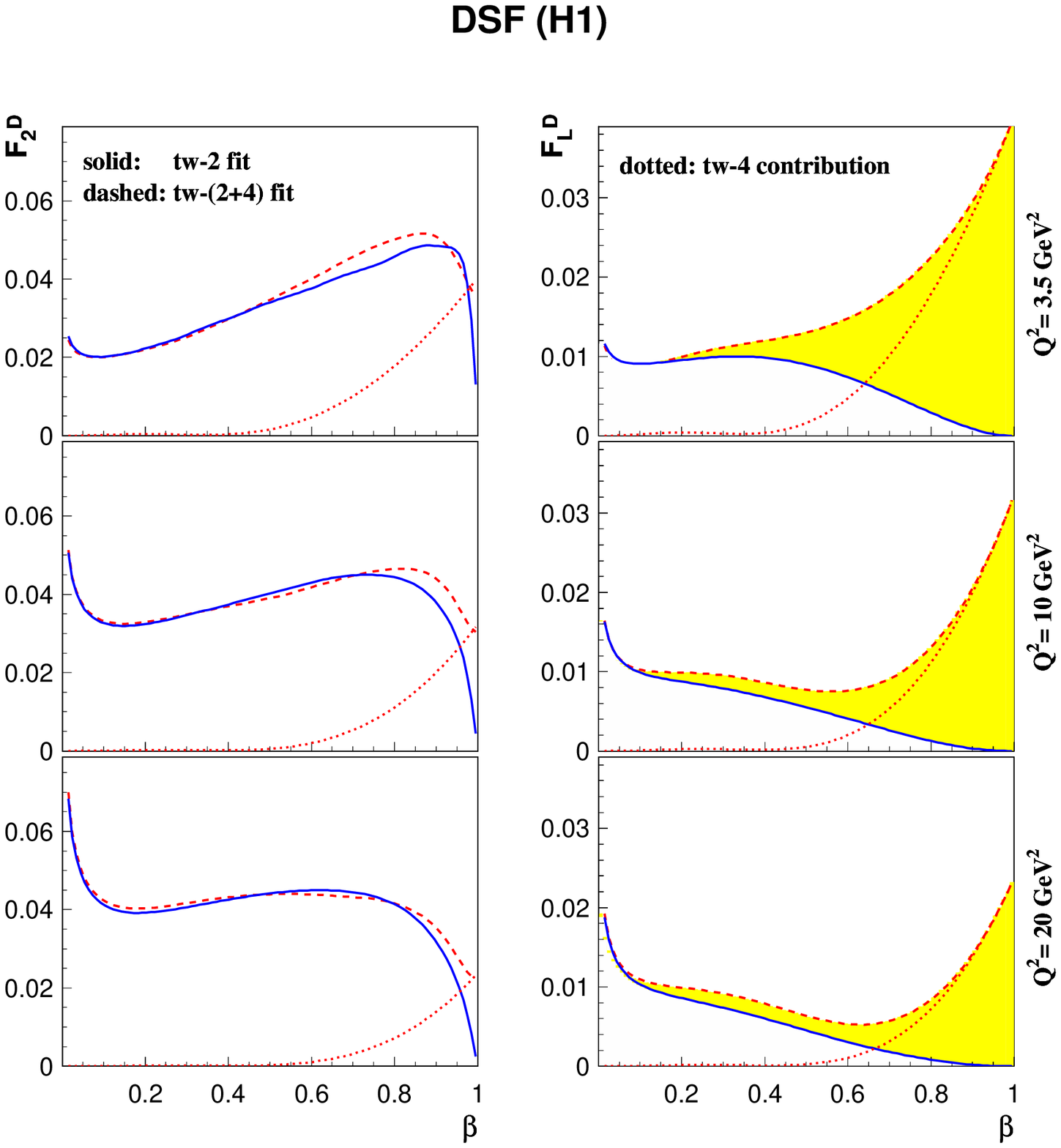,width=17cm}
\caption{Diffractive structure functions $F_2^{D(3)}$ (left) and $F_L^{D(3)}$ (right) from fits to H1 data for $\xp=10^{-3}$.
The  band shows the effect of twist--4 on the predictions for $F_L^{D(3)}$.}
\label{fig:4}
\end{center}
\end{figure}

In Fig.~\ref{fig:4} we present our most important results.
On the left panel, the $F_2^D$ structure function  is shown from both fits, with and without 
the twist--4 contribution (shown by  the dotted lines).  We see no 
significant difference between these two results. 
However,  the longitudinal structure function  $F_L^D$ differs significantly for the two fits (right panel) due to the  twist--4 contribution.
Let us emphasise that both sets of curves were
found in the fits which  well describe the existing data on $\sigma_r^D$, including the large
$\beta$ region. 
Thus, an independent {\it measurement} of $F_L^D$ in this region would be an
important test of the QCD mechanism of diffraction. 

\subsection{ZEUS  data}

The results of same fits performed for the ZEUS data are shown in the last two rows of Table~\ref{tabela:2}. This time the Regge contribution (\ref{eq:F_2R}) is not necessary since fits give the reggeon normalisation $A_R\approx 0$. In general, the fit quality is worse than for the H1 data.

\begin{figure}[p]
\begin{center}
\psfig{figure=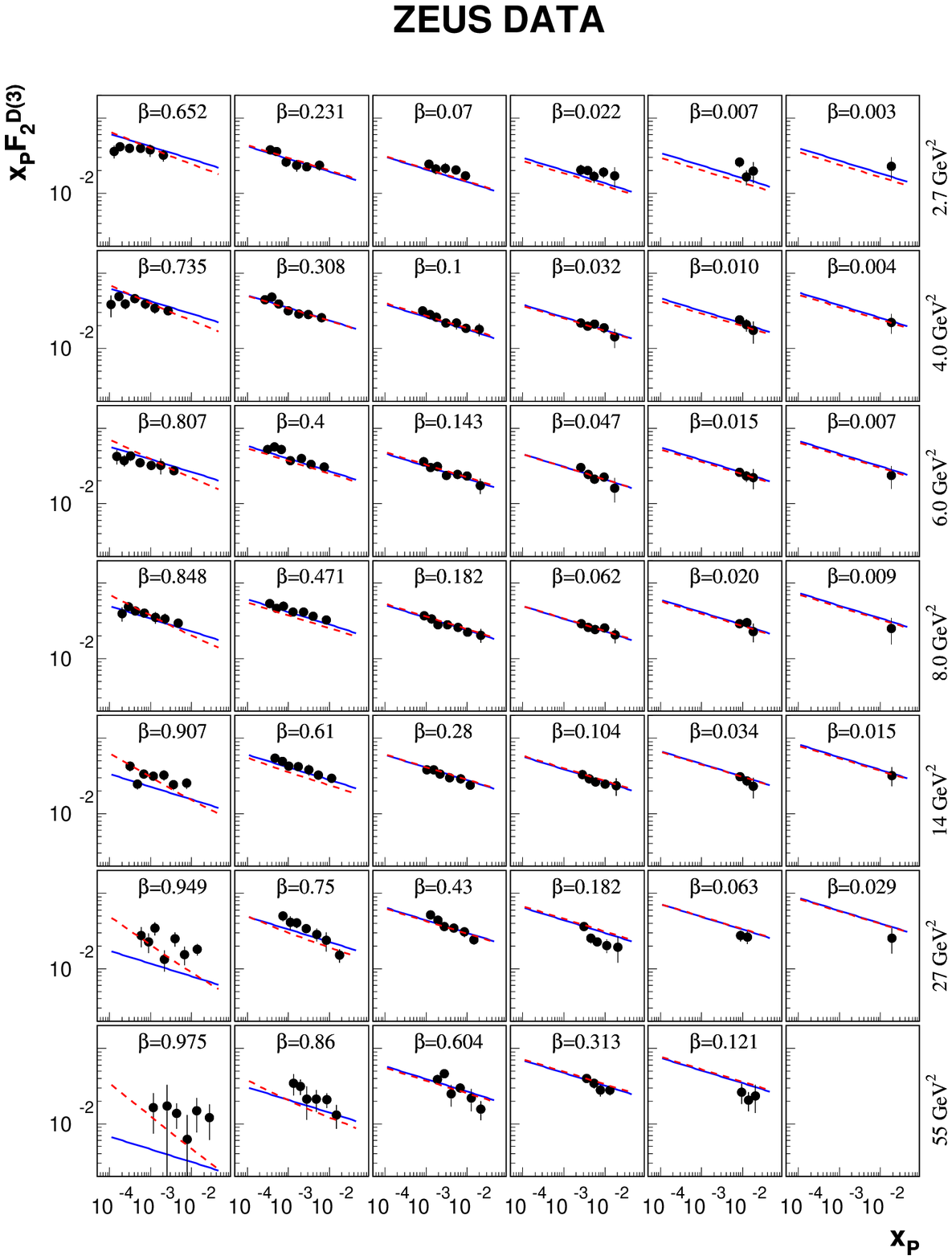,width=17cm}
\caption{Diffractive structure function $F_2^{D(3)}$ as a function $\xp$ for ZEUS data. Solid lines: twist--2 fit, dashed lines: twist--(2+4) fit.}
\label{fig:5}
\end{center}
\end{figure}

\begin{figure}[p]
\begin{center}
\psfig{figure=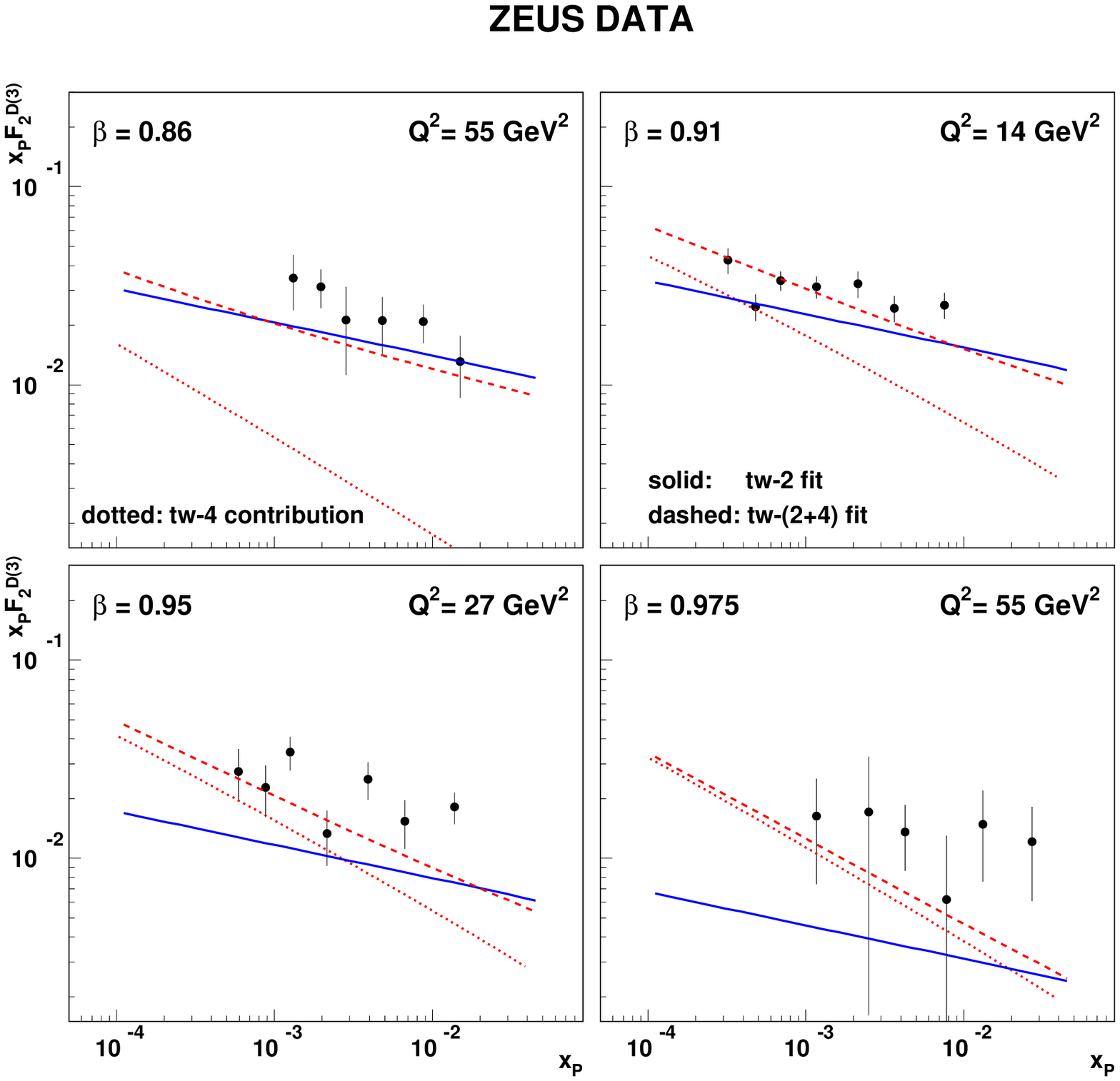,width=17cm}
\caption{Diffractive structure function  $F_2^{D(3)}$ for ZEUS data at
large values of $\beta$ against fit curves.}
\label{fig:6}
\end{center}
\end{figure}

As shown in Fig.~\ref{fig:5},
the biggest difference between the twist--2 and twist--(2+4) results occurs at large $\beta$
values. This is analysed  in detail in Fig.~\ref{fig:6}. We see that the presence of
the twist--4 term in the fit (dashed lines) improves the agreement with the data in this region.
In particular, a steep dependence of $F_2^D$ on $\xp$ is better reproduced by the
twist--(2+4) fit then by the twist--2 one (solid lines).  This dependence is  to large extend driven by the twist--4 contribution (dotted lines).

\begin{figure}[p]
\begin{center}
\psfig{figure=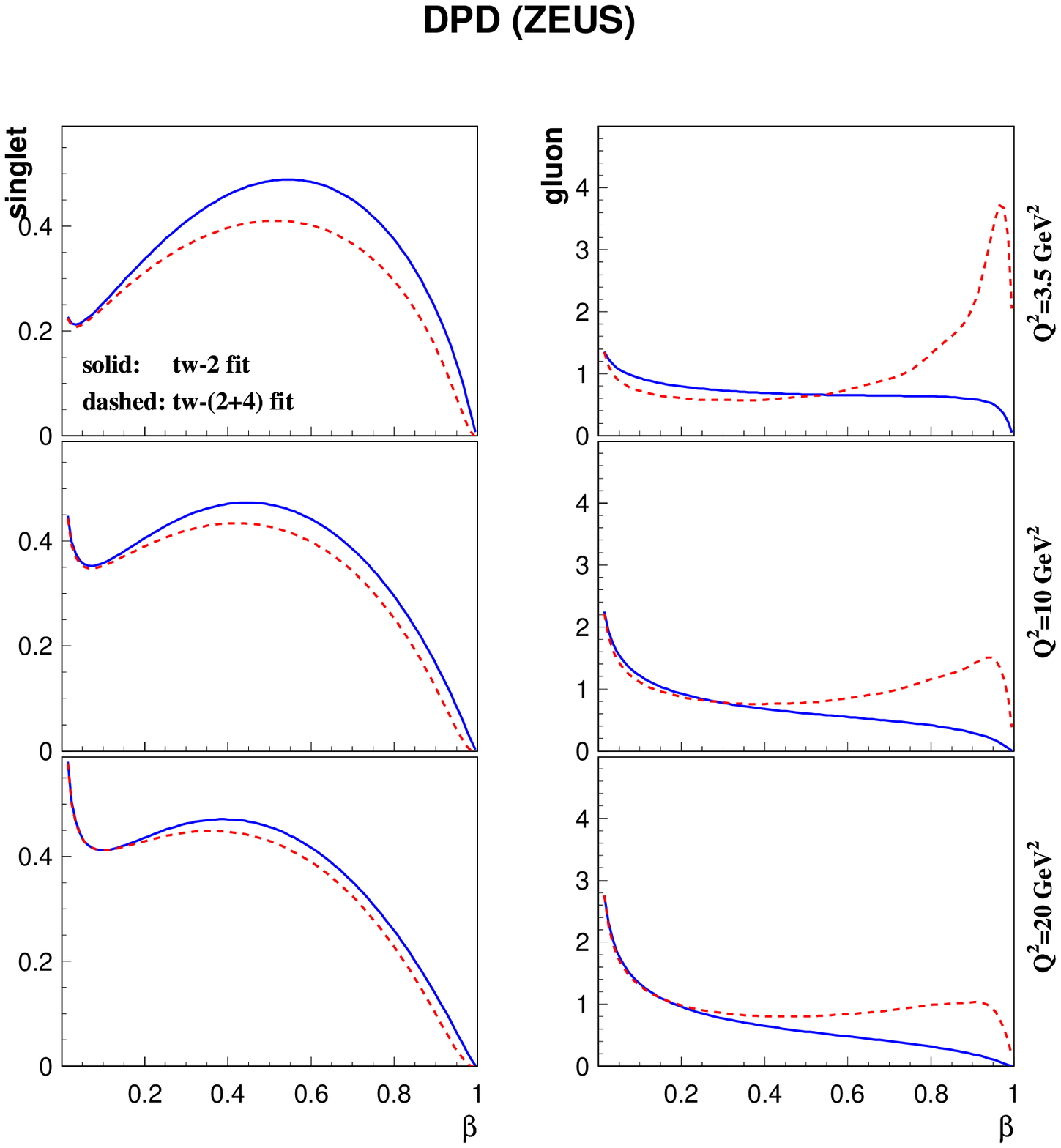,width=17cm}
\caption{Pomeron parton distributions $\beta\Sigma_{\funp}(\beta,Q^2)$ (left) and  
$\beta g_{\funp}(\beta,Q^2)$ (right) from fits to ZEUS data.}
\label{fig:8}
\end{center}
\end{figure}

\begin{figure}[p]
\begin{center}
\psfig{figure=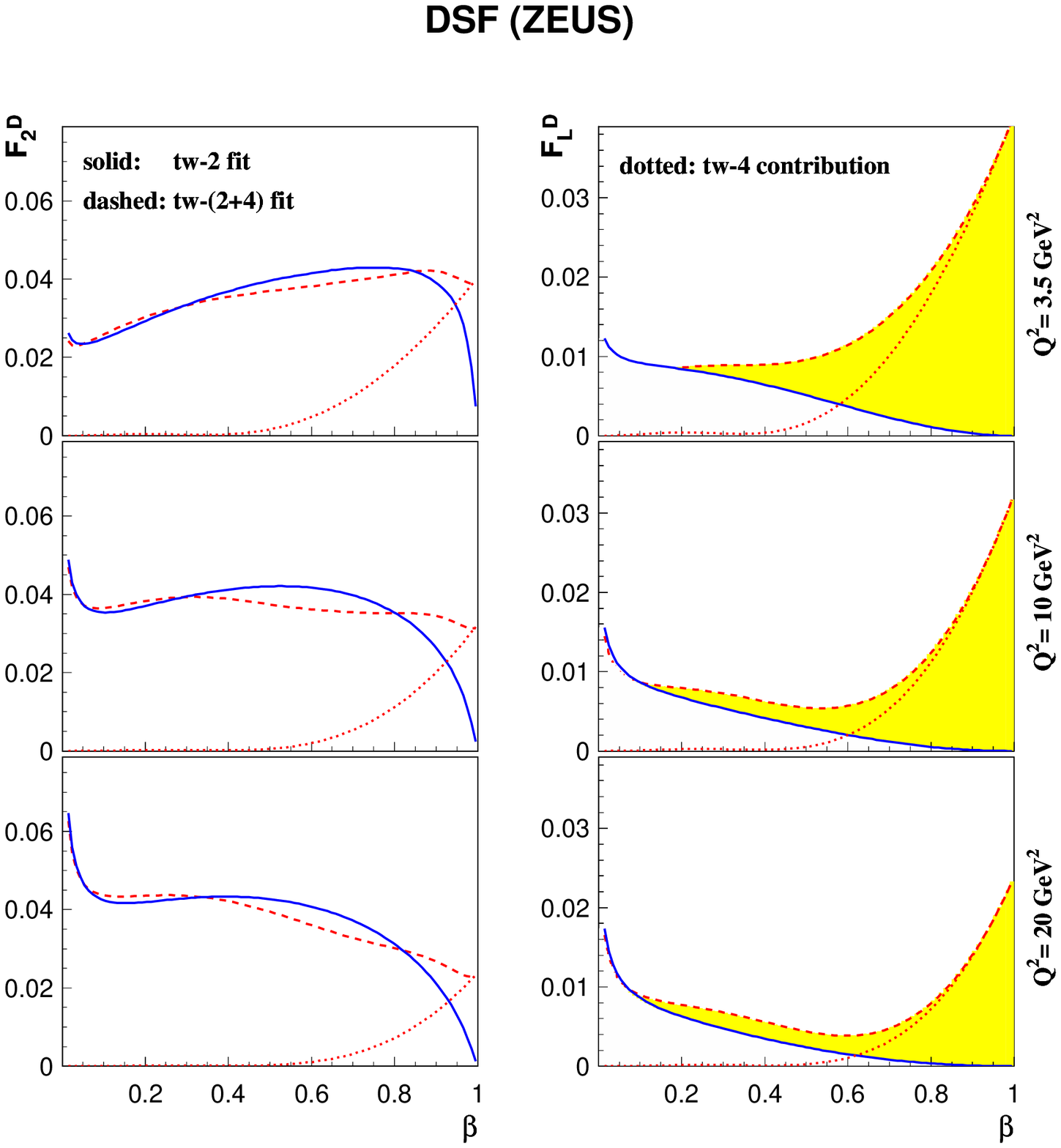,width=17cm}
\caption{Diffractive structure functions  $F_2^{D(3)}$ (left) and $F_L^{D(3)}$ (right) from fits to ZEUS data for $\xp=10^{-3}$. The band shows the effect of twist--4 on the predictions for $F_L^{D(3)}$.}
\label{fig:9}
\end{center}
\end{figure}

The behaviour of the diffractive parton distributions and structure functions,
shown in  Figs.~\ref{fig:8} and \ref{fig:9}, respectively, is very similar to that
found for the H1 data. The gluon distribution from the fit with twist--4
is stronger peaked near $\beta\approx 1$ and the longitudinal structure functions 
in the large $\beta$ region is dominated by the twist--4 contribution.

\begin{figure}[p]
\begin{center}
\psfig{figure=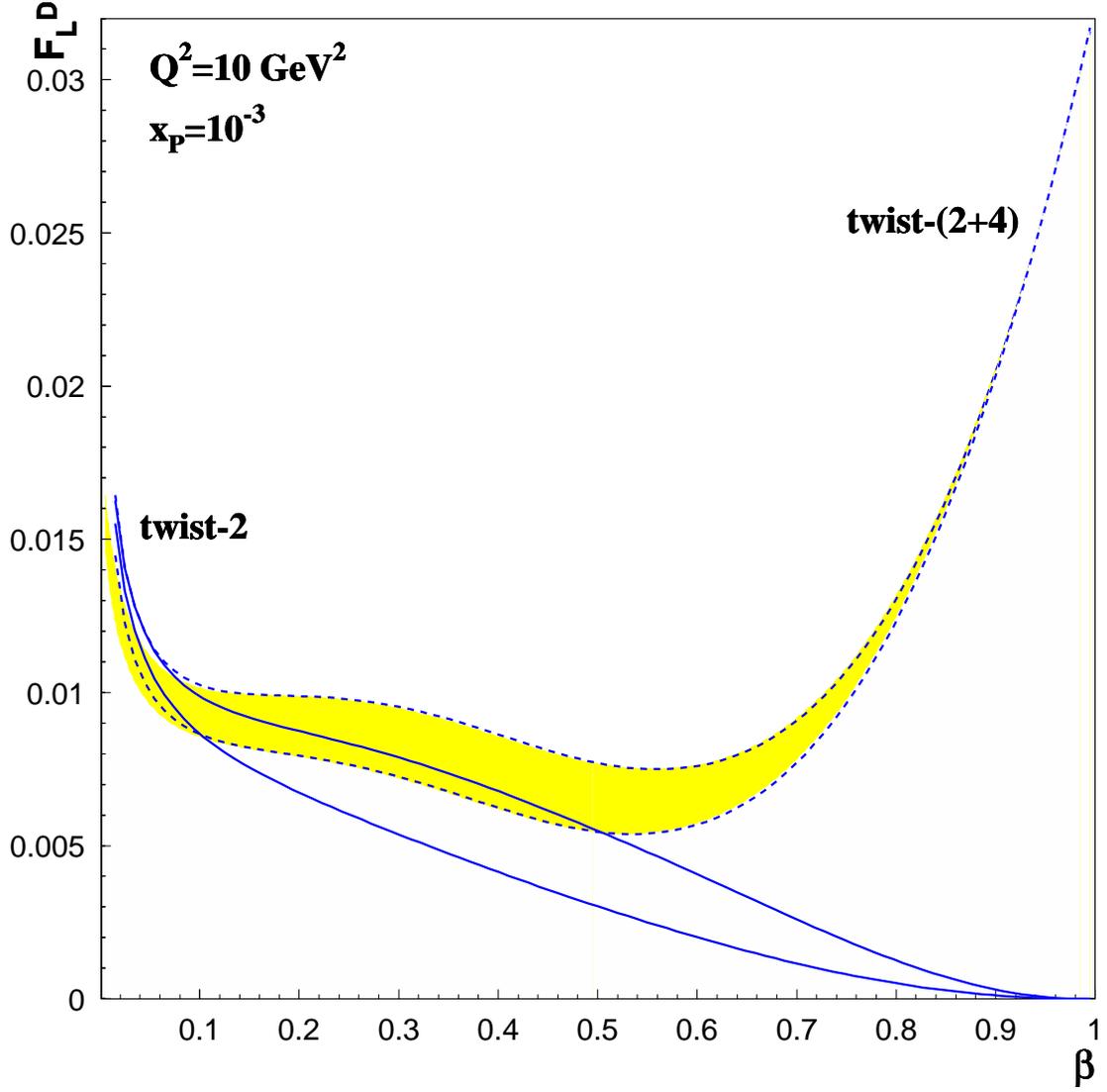,width=17cm}
\caption{Predictions for $F_L^{D(3)}$ for $\xp=10^{-3}$ and $Q^2=10~{\rm GeV^2}$ 
from the twist--(2+4) fits to the H1 (upper dashed line) and ZEUS (lower dashed line) data. The solid lines show predictions from pure twist--2 fits to the H1 (upper) and ZEUS (lower) data.}
\label{fig:10}
\end{center}
\end{figure}

We summarise the effect of the twist--4 contribution in Fig.~\ref{fig:10} showing
the predictions for the longitudinal diffractive structure function $F_L^D$.
Ignoring this contribution, we find the two solid curves
coming from the pure twist--2 analysis of the H1 (upper) and ZEUS (lower) data.
With twist--4, the dashed curves are found, the upper one from the H1 data and the lower one
from the ZEUS data. There is a significant difference between these two predictions in the region of large $\beta$. We believe that the effect of the twist--4 contribution will be confirmed by the forthcoming analysis of the HERA data.


\section{Conclusions}
\label{sec:6}

We performed fits of the  diffractive parton distributions to new diffractive data 
from the H1 and ZEUS collaborations at HERA. In addition to the standard twist--2 formulae, we also considered the  twist--4 contribution which dominates in the region of large $\beta$. This contribution comes from the diffractive production of the $q\qbar$ pair
by the longitudinally polarised virtual photons.
The effect of the twist--4 contribution on the diffractive parton distributions
and structure functions was carefully examined. 
The twist--4 contribution leads to the gluon distribution which is 
peaked  stronger at $\beta\approx 1$ than in the case without twist--4.

The main result of our analysis is a new prediction for 
the longitudinal diffractive structure function $F_L^D$. The twist--4 term 
significantly enhances $F_L^D$ in the region of large $\beta$.
A measurement of this function at HERA in the region of large $\beta$ should confirm the presented expectations which are based on the perturbative QCD calculations. 
The obtained diffractive parton distributions
can also be used in the analysis of  diffractive processes at the LHC,
in particular, to the estimation of the background to the diffractive Higgs production, see \cite{Khoze:2007hx} for a recent discussion.

\vskip 1cm
\centerline{{\bf Acknowledgements}}
This research has been partly supported by MEiN research grant~1~P03B~028~28 (2005-08)
and by EU grant HEPTOOLS, MRTN-CT-2006-035505. The hospitality of the 
Galileo Galilei Institute for Theoretical Physics in Florence and partial support
of INFN during the completion of this work is gratefully acknowledged. 

\vskip 2cm

\bibliographystyle{h-physrev4}
\bibliography{mybib}

\newpage

\newpage

\newpage

\newpage

\newpage

\newpage

\newpage

\newpage

\newpage

\newpage

\end{document}